# Reinforcement Learning for Resource Provisioning in Vehicular Cloud

Mohammad A. Salahuddin[†], *Member, IEEE*, Ala Al-Fuqaha, *Senior Member, IEEE*, and Mohsen Guizani, *Fellow, IEEE*

*Abstract*—This article presents a concise view of vehicular clouds that incorporates various vehicular cloud models, which have been proposed, to date. Essentially, they all extend the traditional cloud and its utility computing functionalities across the entities in the vehicular ad hoc network (VANET). These entities include fixed road-side units (RSUs), on-board units (OBUs) embedded in the vehicle and personal smart devices of the driver and passengers. Cumulatively, these entities yield abundant processing, storage, sensing and communication resources. However, vehicular clouds require novel resource provisioning techniques, which can address the intrinsic challenges of (i) dynamic demands for the resources and (ii) stringent QoS requirements. In this article, we show the benefits of reinforcement learning based techniques for resource provisioning in the vehicular cloud. The learning techniques can perceive long term benefits and are ideal for minimizing the overhead of resource provisioning for vehicular clouds.

## I. INTRODUCTION

IN the absence of standardization in vehicular clouds, the term vehicular clouds, itself, is ambiguous. It has been used to define various different architectures that essentially extend cloud computing functionalities to Vehicular Ad hoc Networks (VANETs). VANETs consist of static infrastructure road-side units (RSUs), and mobile and stationary vehicles with on-board units (OBUs).

Typically, the RSUs communicate via disparate communication media, such as, wired Ethernet, Fiber Optic communication or wireless 3G/4G cellular network channels. Whereas, the vehicles communicate amongst each other, vehicle-to-vehicle (V2V), and with the infrastructure, vehicle-to-infrastructure (V2I) using IEEE 802.11p Wireless Access for Vehicular Environment (WAVE), which is specifically adapted for high-speed data transfer in transient wireless communication channels.

Intelligent Transportation Systems (ITS) leverage VANET to offer services primarily for road safety, such as, lane departure, vehicle proximity indicator, etc. However, popular ITS services also include efficiency, convenience and infotainment services, such as, live traffic congestion reports, on-demand multimedia video streaming, etc. In support of these resource-intensive services, it was proposed that RSUs act as gateways to the cloud, where resource intensive tasks can be offloaded [1]. This cloud-like computing was also extended to include OBUs and RSUs [1] and to the personal devices of drivers and passengers [2]. Generally, it is agreed that vehicular clouds are RSU and OBU resources that provide a pool of processing, sensing, storage and communication resources that are dynamically provisioned for ITS services.

Inherently, the vehicular cloud is a harsh environment with hard and soft QoS requirements on its services [3]. It has intermittent communication links that require high-speed data transfer in a mobile topology, with dynamically changing demands for services with varying QoS requirements. Therefore, efficient and dynamic resource provisioning is crucial for the success of vehicular clouds.

A naïve approach to resource provisioning is static resource allocation, where ITS service providers estimate the demand for their service through statistical analysis and market research. They infer the vehicular cloud resources required to satisfactorily meet the service demands and lease those resources in the vehicular cloud for a given time period. However, static resource allocation will rarely ever be exact. On one hand, under-provisioning will compromise service performance and result in dissatisfied service consumers. On the other hand, over-provisioning results in idle time of leased resources, increasing operating costs and reducing profitability.

A dynamic resource allocation technique will be more reactive to the demand for a service. It will incur lower under– and, or over–provisioning of resources, as it will be closer to the true demand for the service. As the demand for services change over time, the resource provisioning techniques will elastically adapt to the changes and increase or decrease resources allocated for the service(s).

However, continuous and dynamic adaptation of resource provisioning incurs overhead, associated with re-processing the types of resources required, amount of resources required, and the placement of those resources in the vehicular cloud.

M. A. Salahuddin is with Department of Computer Science, Université du Québec à Montréal, Montreal, Quebec, Canada (e-mail: mohammad.salahuddin@ieee.org).

A. Al-Fuqaha is with Department of Computer Science, Western Michigan University, Kalamazoo, Michigan, USA (e-mail: ala@ieee.org).

M. Guizani is with Computer Science and Engineering, Qatar University, Doha, Qatar (e-mail: mguizani@ieee.org).

[†]The majority of this work was done when the author was affiliated with Western Michigan University, Kalamazoo, Michigan, USA.

Therefore, the objectives of dynamic resource provisioning are three-fold, (a) minimize cost of resource provisioning, (b) maximize QoS for end-user perceived latency, and (c) minimize dynamic resource provisioning overhead.

Short of an efficient resource allocation technique, the vehicular clouds will be limited in their functionality and benefits. Reinforcement learning techniques can be exploited and are undoubtedly crucial for designing efficient dynamic resource provisioning heuristics for vehicular clouds. Specifically, Markov Decision Process (MDP), a reinforcement based learning technique, is ideal for dynamic resource allocation. Inherently, its decision making process maximizes the long term benefit. It overcomes the limitations of heuristics that have a myopic approach in decision making. Consequentially, the dynamic resource allocation overhead is minimized without sacrificing QoS for vehicular cloud users.

In this article, our contributions can be delineated as follows. First, we demystify vehicular clouds and present a concise vehicular cloud model that integrates various proposed vehicular cloud models in the recent literature. Second, we discuss the services offered as utilities in vehicular clouds. Third, we present challenges of resource allocation in vehicular clouds and show how they can benefit from reinforcement learning based, Markov Decision Process (MDP). Lastly, we show how MDP can benefit resource allocation by minimizing reallocation overhead in vehicular clouds.

## II. OVERVIEW OF VEHICULAR CLOUDS

The traditional cloud infrastructure consists of geographically dispersed large-scale data centers with hundreds and thousands of machines available to the public. The integration of this cloud infrastructure and its computing paradigm with VANET, gave rise to various different vehicular clouds (VCs). Broadly speaking, the vision is to harness the abundant resources in the OBUs, RSUs and integrating them with the seemingly infinite resources of the traditional cloud data centers, [1], [2], [3], [4], [5], just to mention a few. Initially, a VC was a temporary cloud of OBU resources. This temporary cloud could be static, consisting of a fleet of stationary, i.e. parked vehicles or dynamic, with slow or fast moving vehicles, and, or their hybrid. Eventually, VCs began to incorporate RSU resources and progressed to extend RSUs as gateways to the traditional cloud. We illustrate a classification of vehicular cloud models, in Fig. 1.

We integrate these different visions and models concisely, into a vehicular cloud that seamlessly and transparently integrates heterogeneous devices, their resources and the disparate communication media, as illustrated in Fig. 2. Essentially, the vehicular cloud, as in Fig. 2, is demystified and abstracted as a network of nodes and edges and the set of services they offer as utilities. Service providers lease the vehicular cloud services to build and offer ITS safety, efficiency, and convenience and infotainment services to ITS users.

### A. Network of Nodes and Edges

The vehicular cloud consists of a network of nodes and edges connecting the nodes. The vehicular cloud nodes can be broadly classified as PANs of smart devices of driver and passenger, OBUs embedded in the vehicles, fixed RSUs, and the traditional cloud data centers.

As illustrated in Fig. 2, the vehicular cloud nodes include but are not limited to those physical devices that offer sensory, storage, computation and communication resources. The sensory nodes consist of sensors instilled in the vehicles, fixed roadside infrastructure and in-building infrastructure, e.g. microwave radar, acoustic, inductive and photoelectric sensors in the roadside infrastructure, inertial measurement units and photoelectric sensors in vehicle OBUs and surveillance cameras in public buildings, such as public libraries. Similarly, the vehicular cloud has myriad storage and computing resources to boost. It includes the computing resources of myriad OBUs, computing, sensory and storage resources of personal devices of driver and passengers and the infinite computing and storage resources of the traditional cloud data centers.

The edges connecting these vehicular cloud nodes consist of disparate communication links. These communication links are composed of physically wired or wireless communication media operating under different communication protocols that dictate and manage data exchange. For example, the personal devices within a vehicle create a personal area network (PAN) with dedicated short-range communication protocols, such as Bluetooth.

On the other hand, the OBUs in vehicles communicate with each other and fixed RSUs via IEEE 802.11p Wireless Access for Vehicular Environments (WAVE) in vehicle-to-vehicle (V2V) and vehicle-to-infrastructure (V2I) communication, which is specifically designed for high-speed data exchange and handoff in mobile environments. The IEEE 802.11p incorporates the Enhanced Distributed Channel Access (EDCA) mechanism to prioritize messages for safety and non-safety applications, therefore ensuring that safety applications are not compromised.

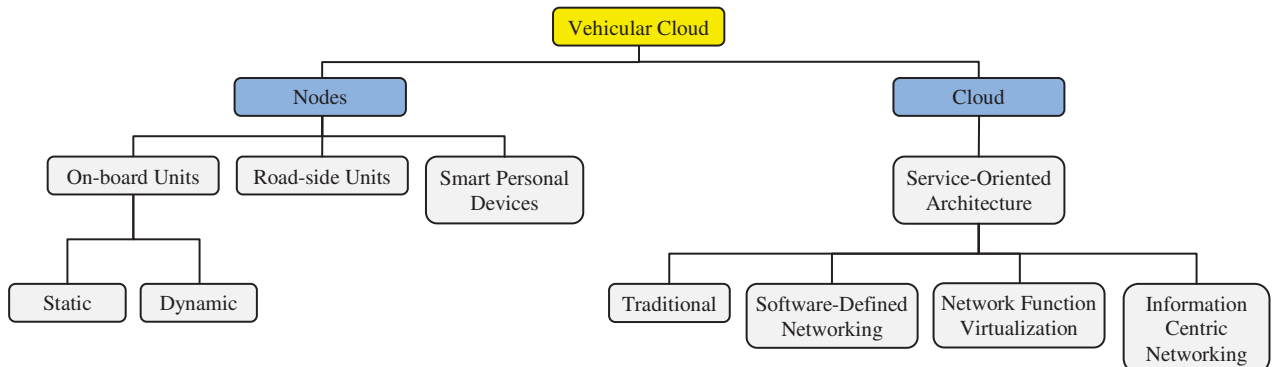

Fig. 1. Classification of vehicular cloud models.

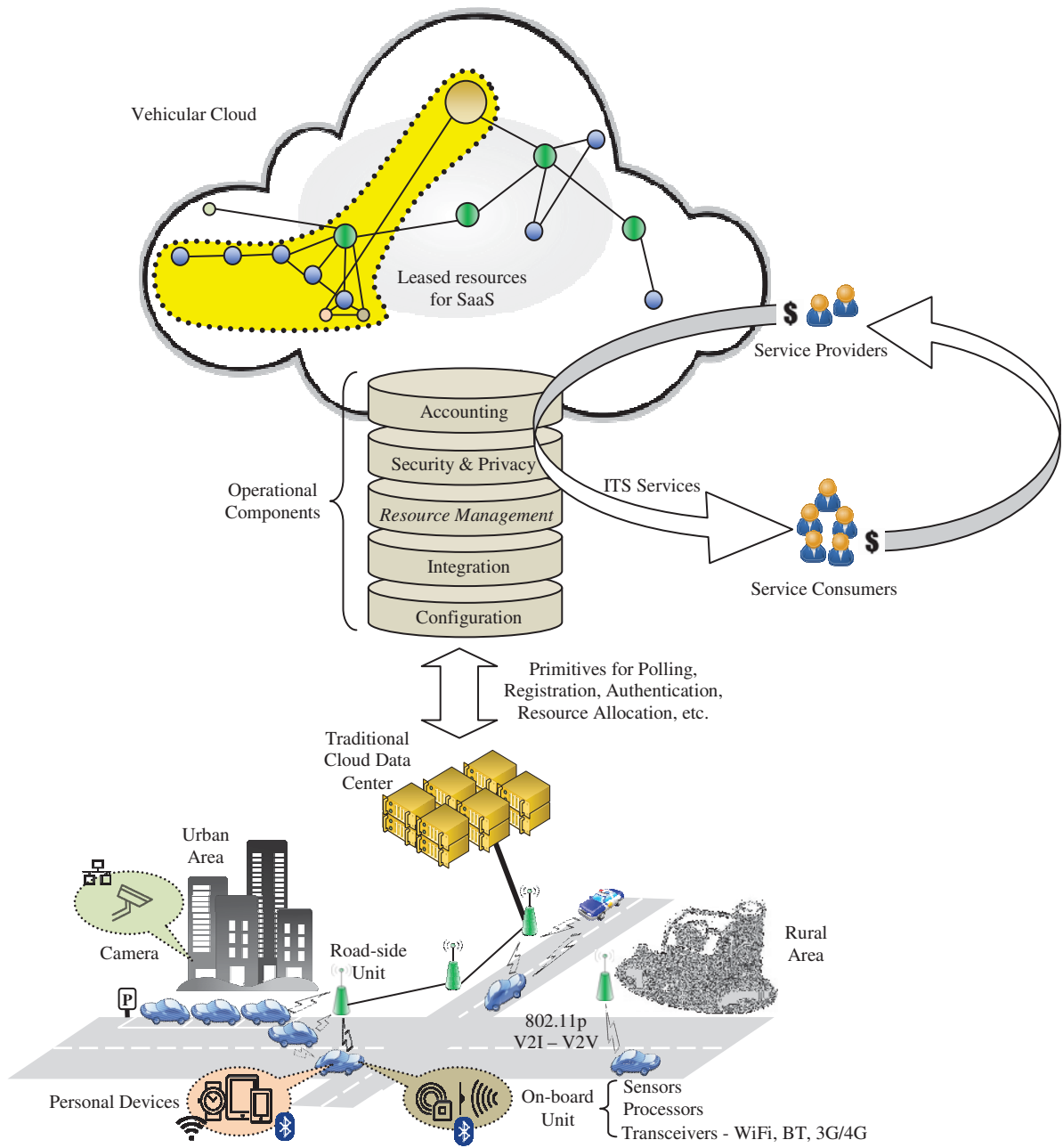

Fig. 2. The vehicular cloud, its nodes, edges & services, its operational components and interaction with service providers and consumers.

The fixed RSUs and other road-side infrastructure are connected via reliable wired high-speed Fiber Optic, Ethernet cables, or wireless cellular 3G/4G networks. Finally, the gateway to the traditional cloud and the inter-cloud communication is via the Internet. The gateway nodes can include RSU nodes [3] and, or OBUs [5]. Revolutionary internetworking technologies, such as, Software-Defined Networking (SDN) [6] and Network Function Virtualization (NFV) [7], enable elasticity and programmability of the vehicular cloud [3]. Whereas, [4] and [8] leverage Information and Content Centric Networking (ICN/CCN) and Service-Oriented Architecture (SOA), and context-aware technologies, respectively, for building efficient vehicular cloud models.

### B. Services – Software-as-a-Service (SaaS)

The vehicular cloud offers Software-as-a-Service (SaaS) as utilities. Fundamentally, this implies that the hardware, system software, its applications and the bandwidth of the communication channels, of the personal smart devices, OBUs, RSUs, and traditional cloud data centers are resources sold to the public as pay-as-you-go services.

It is important to realize that essentially any vehicular cloud service can be built atop SaaS and offered through utility computing, as a utility. For example, if a vehicular cloud was offering Platform-as-a-Service (PaaS), which it formally defined as a convenience service of live traffic reports, then it will use the SaaS of the vehicular cloud to build a *platform*. This platform includes the storage, processing and communication bandwidth resources of the mobile OBUs in the vehicular cloud. Therefore, the PaaS will consume OBU bandwidth for data transfer, storage space on the disk of OBU for the application and its data, and the CPU cycles of the OBU for processing and integrating the data pertaining to location and traffic conditions for its PaaS service. In this way, the SaaS services of the vehicular cloud will be utilities for the PaaS service providers. And the PaaS will be an ITS service utility for the service consumers as in Fig. 2.

*C. Operational Components*

Evidently, there will be software components for building, maintaining and managing the vehicular cloud. They can be broadly classified as network configuration, integration, resource management, quality assurance, threat detection and privacy and accounting.

The network configuration component maintains the network topology that is constantly changing due to mobile vehicles entering and leaving the network. The integration component seamlessly integrates the heterogeneous hardware and communication channels for a seemingly transparent network of nodes and edges. The fundamental resources of the network are managed by a resource management component that allocates, reclaims and reallocates resources across the network for various SaaS services. Essentially, the resources of the vehicular cloud include the processing, storage, sensing and communication capabilities of the entities (OBUs, RSUs, and PANs, traditional cloud) in the vehicular cloud.

Inevitably, the harbinger of success for vehicular clouds lies in end-user perceived QoS and resilience in face of privacy and security breach. The QoS for end-users will be monitored and maintained by the quality assurance component. The threat detection, prevention and perseverance component prevails in analyzing security and privacy vulnerabilities and instilling resilience in the network. Finally, the accounting component will be undoubtedly, the financial, statistical and analytical software required for monitoring and recording the SaaS utilities and billing the vehicular cloud users, that is, ITS service providers and users.

III. PROVISIONING IN RESOURCE MANAGEMENT COMPONENT

The fundamental characteristic of the vehicular cloud is the abundant resource utilities. Providers of ITS services, such as safety and infotainment, lease SaaS from the vehicular cloud. There are various challenges in leasing these resources.

*A. Challenges in Resource Management*

Typically, the SaaS utilities can be decomposed into communication resources leased for data collection, and processing and storage resources leased for application execution and data storage. These utilities are offered with the Service Layer Agreement (SLA) between the vehicular cloud provider and the ITS service providers.

The SLA delineates the terms and conditions for the service(s), including QoS factors, such as, service uptime, service downtime, service availability, etc. The QoS factors determine the resources that are allocated to the service. For example, the volume of requests for the service and the hard or soft QoS thresholds are used to deduce the amount of resources required. However, the resources required change continuously. This can be simply attributed to changes in demand for the service(s), or non-trivially to the underlying topology. The collective resources of the vehicular cloud changes as mobile nodes enter or leave the area. Therefore, the available resources change. Thus, the resources required for service(s) also change.

The *actual* resources required for satisfying the demands for the service are unknown until after-the-fact. However, for uninterrupted service, the resources must be allocated a priori. Therefore, consequential and fundamental for SLA and QoS is cost efficient and low latency resource management.

In particular, we focus on resource provisioning. For efficient resource provisioning, the demands are critically analyzed to deduce the type of resources required, quantity of each type of resource required and the placement of the resources in the vehicular cloud. This resource provisioning problem in vehicular cloud is analogous to resource allocation problem in cloud computing ([9], [10]), which has been shown to belong to the class of computational intensive, non-polynomial time solution problems, known as NP-Hard problems. Furthermore, due to the transient nature of vehicular cloud, its varying QoS requirements and high spectrum of demands, efficient resource provisioning becomes even more challenging.

Therefore, large-scale resource provisioning problems are deemed intractable with respect to optimality. However, efficient suboptimal heuristics can be designed for resource provisioning in vehicular clouds, such that, they minimize cost of utilities and maximize QoS for users. Broadly speaking, resource provisioning heuristics can be decomposed into static and dynamic resource provisioning, as illustrated in Fig. 3.

In the static resource allocation technique, service providers predetermine the required resources, say $r_2$ in Fig. 3,

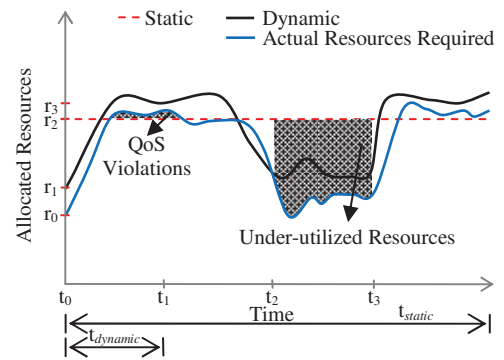

Fig. 3. Resource allocation techniques and their affects.

based on an *expected* demand from end-users over a long period of time, that is, window of time $t_{static}$. Static resource allocation is simple to implement and maintain. However, in terms of resource utilization and QoS, this resource allocation technique is only as good as the estimate. Therefore, if the demand for the service deviates significantly from the estimate, then either the resources will suffer from underutilization or there will be QoS violations.

Underutilized resources, during time interval $t_2 - t_3$, as illustrated in Fig. 3, are the resources allocated under the static resource provisioning technique that are idle, and are economic and financial liabilities. On the other hand, conservative allocations of resources are not suitable for sudden and unexpected surges in service demands. Numerous anecdotes of flash crowds, flash events and SlashDot effects act as reminders of the debilitating consequences of uncalculated resource allocation techniques. These can lead to QoS violations, e.g. at time instant $t_1$ in Fig. 3. The QoS violations can be mild, such that they result in low resolution for multimedia delivering services, or severe, where there is intermittent service delivery or complete service unavailability. QoS degradation for service consumers will lead directly to loss of vehicular cloud users, i.e. service providers and consumers, and its ability to generate revenue. Therefore, appropriately allocating resources for services is critical to their success.

Dynamic resource allocation techniques adapt the resources allocated to meet the change in demands for services. As illustrated in Fig. 3, as the demands change unpredictably over time the resources required to meet those demands also change. To adapt to these changes, dynamic resource allocation techniques estimate the resources required over shorter, when compared to static schemes, time windows, $t_{dynamic}$.

This allows service providers to change the resources allocated for their services, so that the resource allocated more closely follows the actual amount of resources required to meet the demand for the service(s). However, dynamic resource allocation, as illustrated in Fig. 3, is not free from QoS violation or under utilization of resources. Mathematically, the desire is to design dynamic resource allocation techniques, such that, the function that depicts allocated resources is a lower and upper asymptote, bound, on the actual resources required to meet the service demands.

However, the continuous allocation and re-allocation of resources can amount to significant overhead. The overhead pertaining to dynamic resource provisioning includes resources required for moving data for services across the vehicular cloud for those services. For example, network bandwidth resources are allocated and consumed for migrating data pertaining to an efficiency landmark-based parking assistance service from one physical vehicular cloud location to another. Furthermore, new resources have to be instantiated and allocated at the new vehicular cloud location, such as, parallel image processing and storage.

Reinforcement based learning techniques are useful for decision making in dynamic resource provisioning and allocation techniques. Cordeschi *et al.* [11] consider provisioning and scheduling of resources, by considering physical channel and medium access layers as resources. On the other hand, Zhang *et al.* [12] resource provisioning technique considers resource scheduling for interactive multimedia services in mobile clouds, which can be considered as special cases of vehicular clouds.

In this article, we consider that the resources that are allocated for services will be contained in virtual machines (VMs), and hence focus on resource allocation for VMs in vehicular clouds.

### B. Resource Provisioning for VMs in Vehicular Clouds

Traditionally, resource provision is accomplished through virtualization, where VMs are instantaneously activated and deactivated with application specific code. Yu *et al.* [13] use game theoretic approaches for provisioning resources for VMs in the vehicular cloud. Game theoretic approaches are limited by the knowledge of each player, i.e. the VMs that are cooperating in the game. For maximizing the benefits of game theoretic based approaches for resource allocation, it is essential that VMs have complete knowledge of the cost functions of all the other VMs. However, in [3], we apply the Markov Decision Process (MDP), a reinforcement learning technique, to maximize the long term benefit in resource allocation for VMs. MDP techniques are much simpler when compared to game theoretic approaches.

On the other hand, Arkian *et al.* [14] use MDP with Q-learning for selecting the best candidate node for delivering services in a cluster in the vehicular cloud. This selection problem can be abstracted as resource allocation, if vehicle(s) in the cluster are considered as resources. They attribute significant success rates in service delivery to the learning mechanism in the resource allocation techniques. We scrutinized the resource allocation problem [3], from the perspective of resource allocation for VMs using MDP.

### C. Markov Decision Process

Here, we will show benefits of MDP over myopic resource allocation techniques. Consider, the dynamic and actual resource allocation scenarios illustrated in Fig. 3, as the time progresses, the resources allocated for the service(s) change. The resource allocation relies primarily on the demand for the service(s). The changes in the allocated resources, in Fig. 3, are attributed to the change in demand for the service(s). For example, in the time interval $t_0 - t_1$, the demand for services increase, while majority of interval $t_1 - t_2$, sees approximately the same demand. However, the demand for the services decline as time $t_2$ approaches. In time interval $t_2 - t_3$, the service has a constantly lower demand than previous time interval, but it gradually increases by time $t_3$.

Now, consider the network configuration at time $t_0$, as illustrated in Fig. 4. A configuration is a snapshot of the allocated resources, i.e. VMs that are leased in the vehicular cloud to meet QoS for user demands. It is a directory that contains the list of entries. Each 2-tuple entry records the node and the resource leased from that node, in the vehicular cloud. Let us assume that at time $t_0$, an ITS service provider leases resources, say $r_0$ in Fig. 3 and is configured as illustrated in Fig. 2 to meet the demand from ITS service consumers.

At time $t_1$ when the dynamic resource allocation heuristic is provisioning resources for the next time interval $t_1 - t_2$, it infers that service demands have increased from time $t_0$ and re-evaluates the resources required to meet the new demand with QoS, as per SLA requirements. Fig. 4 illustrates two configurations that will meet the increase in demand for the service with QoS. The configurations are different, as they vary in the number of resources allocated, $r_3$ and $r_4$, respectively. The resources allocated, explicitly, includes the number of VMs that are instantiated with application specific code and its data to cater to the demands for the service(s). But, implicitly, it also accounts for overhead pertaining to VM migrations. VM migrations inject traffic into the network for moving the data and code, from one node to another in the vehicular cloud. It is important to note that resources can be abstracted to incorporate resources required for application migration or service replication.

A resource allocation heuristic will deduce a set of configuration(s) and select the best configuration amongst them, based on predetermined criteria. Consider a greedy resource allocation heuristic [3] that selects a configuration, which simply minimizes the allocated resources. Therefore, if we consider that at time $t_1$, as illustrated in Fig. 4, configuration $c_1$ has allocated resources $r_3$, which are lower than the allocated resources $r_4$, of configuration $c_2$, then this heuristic will select configuration $c_1$ as the best configuration. Therefore, as illustrated in Fig. 3, at time $t_1$, the resources allocated are $r_3$.

However, this approach is shortsighted and only maximizes immediate gains. The lack in the learning capability of the heuristic limits it from selecting the configuration that reduces the overhead in the long term. The Markov Decision Process (MDP) is fundamental to overcome this limitation of the myopic heuristics.

The MDP is a discrete time stochastic process, defined by a quad-tuple $<S, A, P, R>$, where $S$ is the set of states and $A$ is set of actions. The transition from state $m$ to state $n$ is based on the action $a \in A$, defined by the probability $P(m, n, a)$, with corresponding reward $R(m, n, a)$. The goal of the MDP is to find a "policy" which maximizes the long term expected reward. The policy delineates the action to be taken in a given state. MDP can be solved using various techniques, such as, Q-learning, policy iteration, value iteration, linear programming, etc. We are interested in policy iteration to get the optimal policy. The policy converges to the optimal policy when maximum reward cannot be improved any further, in successive iterations. The policy dictates the actions to be taken in each state.

We map the MDP for modeling the resource allocation problem presented. The set of states are enumerated to contain all possible configurations, for allocated resources, with varying demands. An action $a \in A$ is defined as the transition from one state to another. The state transitions are deduced, such that all possible valid configurations are connected and the probability of transition from one state to another is equally distributed amongst each valid state. For example, consider the scenario and configurations of Fig. 4. The configurations are analogous to the states, such that $c_0$ $c_1$, $c_2$ are states $c_i \in S \: \forall \: 0 \leq i \leq 2$, with actions $a_j \in A \: \forall \: 1 \leq j \leq 4$. Action $a_1$ is the transition of $c_0$ to $c_1$ and action $a_2$ is the transition of $c_0$ to $c_2$. The transition probabilities are defined as $P(c_0, c_1, a_1) = 0.5$ and $P(c_0, c_2, a_2) = 0.5$. The rewards in $R$ are defined as $R(c_0, c_1, a_1) = maxResources - (r_3 - r_0)$ and $R(c_0, c_2, a_2) = maxResources - (r_4 - r_0)$.

Simplistically, the MDP in this case, will select the policy that dictates state $c_0$ to transition to state $c_2$ on action $a_2$ to maximize reward, which minimizes overhead of resource allocation. At time instant $t_1$, a greedy approach with myopic goals, will select configuration $c_1$, since it requires less resources than configuration $c_2$. However, this gain in resources allocated is short-term, since configuration $c_2$ is required at time $t_2$, and the resource allocation technique incurs an extra overhead for changing the configurations from $c_1$ to $c_2$.

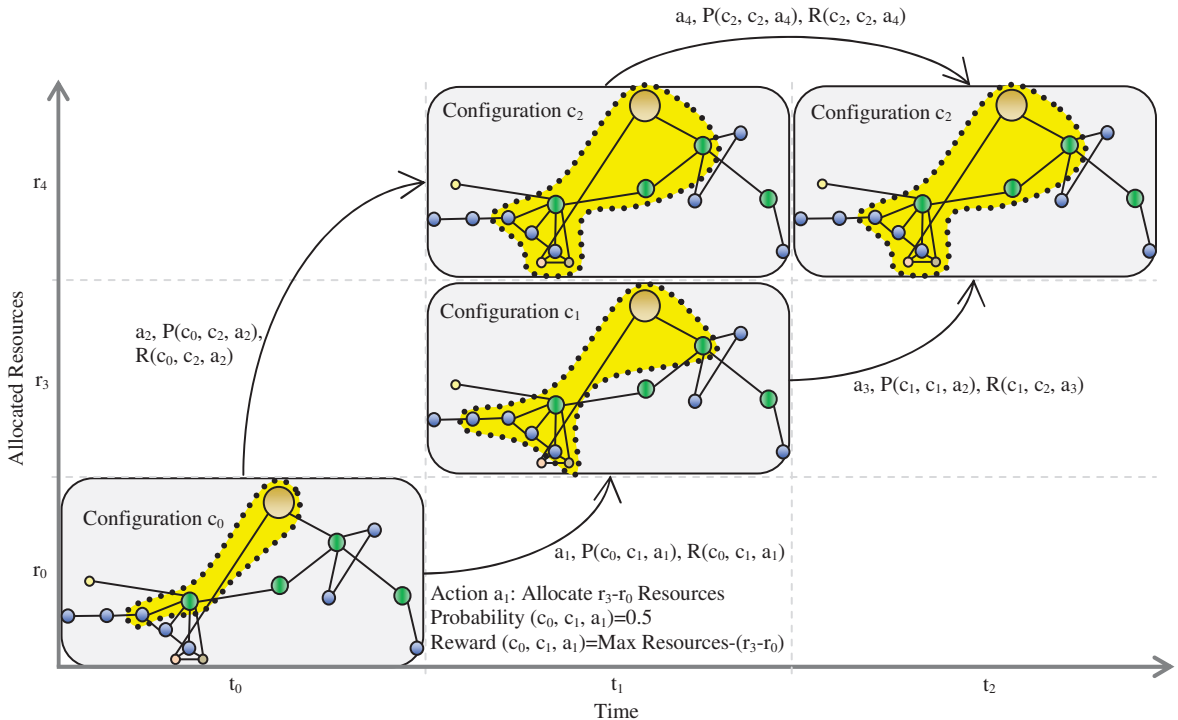

Fig. 4. Markov Decision Process (MDP) for minimizing overhead.

However, for large-scale resource allocation problems, a huge number of states, actions, and their transition probabilities and rewards have to be enumerated and act as input into the MDP. There exist numerous polynomial time algorithms for solving MDP with policy iteration. We performed simulations in MATLAB using a custom reinforcement learning toolbox [15] to illustrate the benefits of MDP over greedy heuristic for resource allocation. For simplicity, we abstracted the resource allocation problem to a static network of RSUs with a fixed amount of available resources, to meet the dynamic demands with QoS, from the users in the vehicular cloud.

The quad-tuple input $<S, A, P, R>$ is deduced, as described and illustrated previously and the MDP is instantiated for the resource allocation problem in vehicular cloud. As illustrated in Fig. 5 and Fig. 6, the policy derived by the MDP will select configurations that in our case minimize the overhead, i.e. VM migrations, over the long term. As illustrated in Fig. 5, a greedy heuristic, e.g. in [3] and MDP may select the same configurations. Therefore, in the worst-case, MDP will perform at least as good as a myopic heuristic.

However, MDP will always optimally select configurations that maximize reward, in the long term. Therefore, similar to Fig. 4, as illustrated in Fig. 6, at time $t_i$ MDP selects a configuration that allocates more resources than the heuristic. However, critical to long-term resource provisioning techniques, MDP overcomes the limitations of the myopic heuristic and allocates lower resources, in the long term, when compared to the greedy heuristic [3]. The long-term resource provisioning in MDP is critical in minimizing resource allocation overhead and total resources allocated in dynamic resource provisioning techniques.

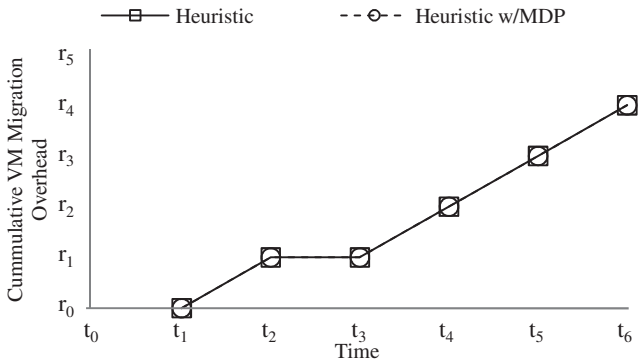

Fig. 5. In the worst-case, MDP will perform as well as a myopic heuristic.

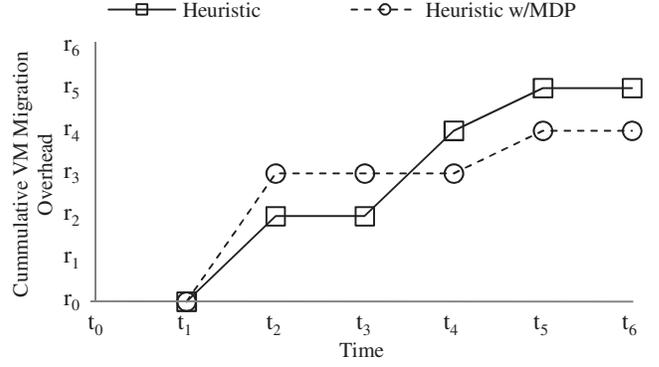

Fig. 6. MDP selects configuration that allocates resources for the long-term benefit in resource allocation, in this case, minimizing cumulative VM migration overhead.

## IV. CONCLUSION

In this article, we concisely integrated the various proposed state-of-the-art vehicular cloud models. The vehicular cloud model seamlessly integrates the resources from the individual devices and systems in the VANET and traditional cloud. The vehicular cloud offers these resources as Software-as-a-Service (SaaS) offerings. The services are utilities for ITS service providers and consumers. We discussed the operational components that build and maintain the vehicular cloud.

Resource Management is a crucial operational component in the vehicular cloud. In this article, we specifically discuss resource provisioning techniques for resource management. Poorly designed resource allocation techniques reduce the economic success of the vehicular clouds, whereas dynamic resource provisioning techniques are most suitable for meeting the dynamically changing demands with QoS. This article shows the benefits of using reinforcement learning based Markov Decision Process (MDP) in resource provisioning. It provisions resources such that the resources allocated over the long term are minimized.

Though various prototypes and experimental vehicular clouds exist, the economic viability for ITS service providers will be crucial in the wide spread deployment of vehicular cloud infrastructure. Major research challenges exist in the operational components of the vehicular cloud, including primitives for reliable data transfer, load-balancing, security, privacy and data dissemination for resource provisioning in vehicular clouds.

ACKNOWLEDGEMENT

This publication was made possible by NPRP grant # [7-1113-1-199] from the Qatar National Research Fund (a member of Qatar Foundation). The statements made herein are solely the responsibility of the authors.

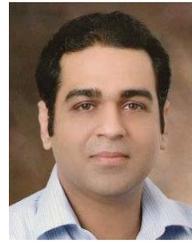

**Mohammad A. Salahuddin** (S'09-M'15) is a Postdoctoral Fellow with the Department of Computer Science, Université du Québec à Montréal. He received his Ph.D. degree in Computer Science from Western Michigan University in 2014. His research interests include Wireless Sensor Networks, QoS/QoE in Vehicular Ad hoc Networks, Internet of Things, Content Delivery Networks, Software-Defined Networks and Cloud Resource Management. He serves as a Technical Program Committee member and reviewer for IEEE journals, magazines and conferences.

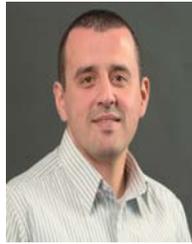

**Ala Al-Fuqaha** (S'00-M'04-SM'09) is currently a Professor and director of NEST Research Lab at the Computer Science Department of Western Michigan University. His research interests include Wireless Vehicular Networks (VANETs), cooperation and spectrum access etiquettes in cognitive radio networks, smart services in support of the Internet of Things, management and planning of software defined networks (SDN). He is currently serving on the editorial board for multiple journals and has served as Technical Program Committee member of many international conferences.

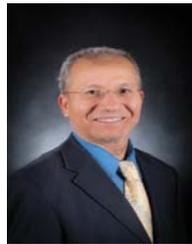

**Mohsen Guizani** (S'85–M'89–SM'99–F'09) received the B.S. (with distinction) and M.S. degrees in electrical engineering, the M.S. and Ph.D. degrees in computer engineering from Syracuse University, Syracuse, NY, USA, in 1984, 1986, 1987, and 1990, respectively. He is currently a Professor and the ECE Department Chair at the University of Idaho, USA. Previously, he served as the Associate Vice President of Graduate Studies and Research, Qatar University, Chair of the Computer Science Department, Western Michigan University, Chair of the Computer Science Department, University of West Florida. He also served in academic positions at the University of Missouri-Kansas City, University of Colorado-Boulder, Syracuse University, and Kuwait University. His research interests include wireless communications and mobile computing, computer networks, mobile cloud computing, security, and smart grid. He currently serves on the editorial boards of several international technical journals and the Founder and the Editor-in-Chief of Wireless Communications and Mobile Computing journal (Wiley). He is the author of nine books and more than 400 publications in refereed journals and conferences. He guest edited a number of special issues in IEEE journals and magazines. He also served as a member, Chair, and the General Chair of a number of international conferences. He was selected as the Best Teaching Assistant for two consecutive years at Syracuse University. He was the Chair of the IEEE Communications Society Wireless Technical Committee and the Chair of the TAOS Technical Committee. He served as the IEEE Computer Society Distinguished Speaker from 2003 to 2005.